\documentclass[]{aa}  

\usepackage{graphicx}
\usepackage{txfonts}
\def\msun{\rm\, {M_\odot}}

\begin{document} 

   \titlerunning{Super-Eddington accretion in quasar hosts}
   \title{Sustained super-Eddington accretion in high-redshift quasars}

   \author{Alessandro Lupi
          \inst{1,}
          \inst{2,}
          \inst{3}\fnmsep\thanks{alessandro.lupi@uninsubria.it}   
          \and
          Giada Quadri\inst{3}
          \and 
          Marta Volonteri
          \inst{4}
          \and
          Monica Colpi
          \inst{3,2}
          \and
          John A. Regan
          \inst{5}
          }

   \institute{
            Dipartimento di Scienza e Alta Tecnologia, Universit\`a degli Studi dell'Insubria, via Valleggio 11, I-22100, Como, Italy
        \and
            INFN, Sezione di Milano-Bicocca, Piazza della Scienza 3, I-20126 Milano, Italy
        \and 
            Dipartimento di Fisica ``G. Occhialini'', Universit\`a degli Studi di Milano-Bicocca, Piazza della Scienza 3, I-20126 Milano, Italy
        \and
            Institut d’Astrophysique de Paris, UMR 7095, 
                CNRS and Sorbonne Universit\'{e}, 98 bis boulevard Arago, 75014 Paris, France
        \and
            Centre for Astrophysics and Space Sciences Maynooth, Department of Theoretical Physics, Maynooth University, Maynooth, Ireland
      }

   \date{Accepted 19/04/2024}

 
  \abstract
   {  Observations of $z \gtrsim 6$ quasars provide information on the early evolution of the most massive black holes (MBHs) and galaxies. Current observations, able to trace both gas and stellar properties, reveal a population of MBHs that is significantly more massive than expected from the local MBH-stellar mass relation. The population lies on, but mostly above, the relation observed in the nearby Universe. This suggests that these objects grew very rapidly. 
     To explain their presence when the Universe was less than 1~Gyr old and to assess the physical conditions for their rapid growth, we explored whether episodes of accretion above the Eddington limit can occur across cosmic epochs. By employing state-of-the-art high-resolution cosmological zoom-in simulations of a $z\sim 7$ quasar, where different accretion regimes are included consistently, together with their associated radiative and kinetic feedback, we show that super-Eddington phases can be sustained for relatively long timescales (tens of millions of years). This allows the MBH to rapidly grow by up to three orders of magnitude, depending on the strength of the kinetic feedback. We also show by means of a semianalytic calculation that the MBH spin remains moderate and does not take on extremely high values during the super-Eddington phases. This results in a lower feedback efficiency, which may allow the rapid growth required to explain over-massive high-redshift MBHs.}



   \maketitle
%

\section{Introduction}

   Massive black holes (MBHs) are ubiquitous in the Universe and inhabit the centre of massive galaxies up to redshift $z\gtrsim 6$ \citep[e.g.][]{fan06,mortlock11,banados18,fan23,maiolino23b}, with masses up to $10^{9-10}\msun$. Observationally, they are commonly identified via gas accretion through the conversion of gravitational energy into radiation, which makes them shine as Active Galactic Nuclei (AGN). They sometimes also produce powerful collimated jets.
   According to  Soltan's argument \citep{soltan82}, the evolution of the AGN luminosity function, and the local MBH-mass density \citep{marconi2004}, MBHs  gain most of their mass via radiatively efficient accretion. Hence, they should have formed from lower-mass black hole `seeds' \citep[see, e.g.][for a review]{Inayoshi20,Volonteri21}.

   In this context, the observations of high-redshift MBHs 
   help placing tight constraints on the minimum initial seed mass, which has to be about $M_{\rm seed}\gtrsim 10^4\msun$ when we assume growth to occur via radiatively efficient accretion with $\eta_{\rm rad}\sim 0.1$ at the Eddington limit. Several groups have studied the formation mechanisms of such `heavy' seeds that facilitate interpreting current observations.  However, the formation of the most massive seeds requires rare peculiar conditions \citep[e.g.][]{latif15,schauer17,lupi21,latif22}. An alternative possibility, which was also suggested by recent results \citep[e.g.][]{davies19,yang23}, is that these MBHs grew rapidly because of intermittent phases of super-Eddington accretion from lighter seeds \citep{madau14,volonteri15,lupi16,pezzulli16}. 
   
    The plausibility of super-Eddington accretion has been demonstrated to occur in the local Universe in tidal disruption events \citep{lin17} and ultra-luminous X-ray sources \citep{bachetti14} and was initially proposed in the context of non-spherical accretion flows within the `slim-disc' model \citep{abramowicz88}. Recent simulations of accretion discs showed that the slim-disc solution, in which the radiation is trapped in the innermost regions of the accretion disc and is advected inwards within the fluid, making the disc moderately luminous, is instead characterised by powerful radiatively driven outflows and jets that escape through a central funnel perpendicular to the disc itself \citep{sadowski16a,sadowski16b,jiang19}.
    
    While most of the studies focussing on accretion above the Eddington rate were performed on the typical scales of the accretion disc, large-scale simulations have commonly neglected this regime and based their results on the common Eddington-limited accretion. Moreover, the inability to resolve the influence radius of the MBH and, in many cases, the low resolution preventing the dense gas from being properly resolved also required additional tuning parameters, such as the $\alpha$ boost factor on the accretion rate \citep{springel05a,dimatteo05} and the coupling efficiency parameter for the accretion-powered feedback (dumped in the form of thermal energy), which was commonly set to values in the range $0.05-0.15$, depending on the numerical technique employed to reproduce local observations \citep{dimatteo05,dubois14}. Even though improved models for accretion and feedback were developed more recently, for instance correcting for the gas angular momentum \citep{tremmel17} or dumping the MBH feedback as radiation-driven winds \citep{choi12,anglesalcazar17FIRE}, the physics of the accretion process above the Eddington limit was almost never considered.
    
    In the last few years, however, several groups have started to account for the possibility of super-Eddington accretion, but often in a very simplistic way. For instance, some works only extended the standard accretion and feedback prescriptions above the Eddington limit \citep[e.g.][]{zhu22,ni22,bhowmick22}, whereas others only considered the radiation trapping and the subsequent decrease in radiative efficiency \citep{lupi16,rennehan23}. A more physically motivated modelling that accounts for the impact of kinetic winds/jets during the super-Eddington phases on the interstellar medium of a galaxy host and on the MBH growth itself has instead only been considered in idealised setups such as in isolated galaxies \citep{massonneau23}, circum-nuclear discs \citep{sassano23}, and atomic cooling haloes just after the formation of a heavy-seed MBH \citep{regan19}.

   In this work, we move forward and study the MBH evolution in the quasar host originally studied in \citet{lupi19b,lupi22}. We properly account for super-Eddington accretion phases in a full cosmological context. The simulation adopts the same initial conditions of the high-redshift quasar originally studied in \citet[][hereafter Paper I and II]{lupi19b,lupi22}. This is paper IV  of a series of papers addressing properties of high-redshift quasar hosts and their MBHs. In paper I, we presented and discussed the main evolution of the target galaxy and its central MBH, focussing on the stellar and gas tracers (total gas and [CII] emission), and found that super-Eddington phases were measured in the simulation, even though accretion was capped at the Eddington limit. In paper II, we extended the analysis by focussing on the dynamics and morphology of the main galaxy as a function of redshift. In paper III (Lupi et al. in prep.) we will discuss the evolution of the entire MBH population that forms during the simulation, and in paper V (Quadri et al. in prep.) we will focus in detail on the impact of the super-Eddington regime on the galaxy host and on the properties of quasar outflows.

   The paper is organised as follows. In Section~\ref{sec:setup}, we recapitulate the setup of the simulation and describe the improvements relative to the previous works. In Section~\ref{sec:results}, we present our results. In Section~\ref{sec:conclusions}, we draw our conclusions.

\section{Numerical setup}
\label{sec:setup}
The simulation followed the evolution of a massive halo ($M_{\rm halo} \sim 3\times 10^{12}\msun$ at $z=6$) that is expected to represent a quasar host \citep{dimatteo17}. The initial conditions were accurately created to match the expected halo mass \citep{dimatteo17,tenneti18} and the galaxy overdensity significance \citep{uchiyama18,mignoli20} via \textsc{music} \citep{hahn13}, adopting the \citet{planck16} cosmological parameters, with $\Omega_{\rm m}=0.3089$, $\Omega_\Lambda = 0.6911$, $\Omega_{\rm b}=0.0489$, $\sigma_8 = 0.8159$, $n_{\rm s} = 0.9667$, and $H_0 = 67.74\,\rm km\, s^{-1} Mpc^{-1}$, without a contribution from radiation and curvature.
From a parent dark-matter-only simulation, we recursively zoomed-in on a Lagrangian volume extending up to 2.5 virial radii of the target halo, following the approach by \citet{fiacconi17} to exclude any contamination by low-resolution dark matter particles within the virial radius.

The simulation was run with \textsc{gizmo} \citep{hopkins15}, a descendant of \textsc{Gadget3} \citep{springel08} and \textsc{Gadget2} \citep{springel05} in its meshless finite-mass mode. 
The spatial resolution of the simulation was set to 40, 10, and 2.5~pc h$^{-1}$ for dark matter, stars, and MBHs, respectively, whereas fully adaptive softening was assumed for the gas component, down to a minimum of $\sim 5$~pc. The mass resolution was $\sim 10^4\msun$ for baryons and $\sim 10^5\msun$ for dark matter.

\subsection{Baryonic physics}
Our simulation was performed with state-of-the-art sub-grid prescriptions that allowed us to follow non-equilibrium chemistry of primordial species, star formation, and stellar feedback in detail,  as well as  MBH seeding, accretion, and feedback. Compared to the previous works of the series, we here slightly revised and improved many of the sub-resolution prescriptions, which we describe below.
\begin{itemize}
    \item Gas thermodynamics and chemistry: We further extended our chemical network to include high-ionisation states of several important species that are commonly observed in quasar hosts, namely C[I-IV], O[I-VI], N[I-V], and Fe[I-II], also accounting for their contribution to the low-temperature cooling of the gas. Since MBHs are commonly surrounded by a hot corona emitting in X-rays, we also incorporated detailed X-ray chemistry calculations in our network, accounting for the impact of Compton heating by the AGN (assuming $T_{\rm Compton}=3.23\times 10^6$~K and $T_{\rm Compton}=8.41\times 10^7$~K for soft and hard X-rays, respectively). This will be discussed in Lupi et al. in preparation in greater detail.
    \item Star formation: We slightly revised our estimate for the turbulent support of the gas as in \citet{hopkins13sf}, accounting for the particle distribution inside the kernel, which gives $\sigma_{\rm turb}=||\nabla v||/5$. Relative to \citet[][L19 hereon]{lupi19b}, we updated our star formation efficiency employing the \cite{padoan12} model, as in \citet{lupi20}.
    \item Stellar mechanical feedback: We redetermined the scalings in \citet{martizzi15} to improve the agreement with their results, that is, we properly accounted for the initial fraction of kinetic and thermal energy during the Sedov-Taylor phase $f_{\rm kin}\sim 0.28$.
    \item Stellar radiative feedback: Instead of the cost-effective approximated radiation transport of L19, we included on-the-fly radiation transport as in \citet{lupi20b}, with the reduced speed of light set to $c_{\rm red}=1000\rm\, km\, s^{-1}$, which is high enough compared to the gas motions to ensure consistent results. In addition, we followed X-ray chemistry, and we therefore now follow radiation in 11 photobins ranging from 0.7 eV up to 10 keV, where 2 bins are used to cover soft (0.2-2 keV) and hard (2-10 keV) X-rays.  
\end{itemize}

\subsection{MBH accretion/feedback and dynamics}
In addition to the changes above, we devised a novel set of prescriptions for MBH growth and dynamics that we discuss in detail in the following. The MBH seeding is instead identical to that in L19 and occurs in galaxies with a stellar mass $>10^8 \msun$ that do not yet host an MBH, that is, galaxies that are identified through an on-the-fly Friends-of-Friends algorithm (see L19 for details). \subsubsection{MBH dynamics}
In most cosmological simulations, the mass and spatial resolution of seed MBHs  is not sufficient to accurately resolve the dynamical friction bringing MBHs to the centre of galaxies, and often also the interaction with other particles, leading to spurious scattering of the MBHs.  For this reason, most simulations include an ad hoc MBH pinning procedure \citep{dimatteo05,schaye15,barai18} that moves the MBH to the potential minimum inside its kernel at every time step. Although effective, this procedure can produce unphysical behaviour such as superluminal motions or artificial suppression of the MBH wandering. A more physically motivated prescription artificially corrects the dynamics accounting for the unresolved dynamical friction effect \citep{dubois13,tremmel15}, as long as the mass ratio of the MBH relative to the other tracers (gas, stars, and dark matter) is high enough \citep{tremmel15,pfister19}. \citet{lupi19b} ensured a reasonable dynamical evolution by seeding the MBH with an already high mass of $M_{\rm BH}=10^6\msun$. In this work, we instead opted for decoupling the MBH mass into a physical mass (used for accretion) and a dynamical mass  (for the dynamics) \citep{anglesalcazar17FIRE}, which evolve together as soon as the first reaches the second. The physical mass was set to $M_{\rm BH}=10^5\msun$, and the dynamical mass $M_{\rm dyn,MBH}$ to a value ten times higher. This ensured a better dynamical evolution. From a physical point of view, this initially higher dynamical MBH mass can be considered as an unresolved stellar envelope (a nuclear stellar cluster) surrounding the MBH, which is commonly found in many galaxies in the local Universe \citep[see][for a review]{neumayer20}. From a numerical point of view, however, it was simply used to avoid the spurious scattering by other particles, which despite the very high resolution adopted in this simulation are still more massive than individual stars.

Unlike \citet{tremmel17} and \citet{ma21}, the unresolved dynamical friction in this work was implemented following the more accurate model by \citet{pfister19}, where the dark matter and stellar distributions are accounted for separately (thus allowing for different distribution functions), and the high-velocity part of the distribution function was also considered. Moreover, we also included gas-driven dynamical friction as described in \citet{tanaka09}, following the prescription in \citet{escala04}.

\subsubsection{MBH accretion and feedback}
One of the main novelties of this work is the inclusion of three accretion regimes covering the entire range of accretion rates. We also account for the different feedback mechanisms relevant in each of them.\footnote{During the drafting  of this work, \citet{rennehan23} also proposed inclusion of the three MBH accretion regimes in simulations, but not the impact of jets during the super-Eddington stage.} We classify them in terms of the Eddington ratio $\lambda=\dot{M}_{\rm BH}/\dot{M}_{\rm Edd}$, where $\dot{M}_{\rm BH}$ is the accretion rate, $\dot{M}_{\rm Edd}=16\, L_{\rm Edd}/c^2$\citep{madau14}, $L_{\rm Edd}$ is the Eddington luminosity, and $c$ is the speed of light. In principle, the radiative efficiency depends on the MBH spin, which in the current simulation is not evolved over time. We set the spin magnitude to a constant value of $a=0.7$, which gives the commonly adopted radiative efficiency in the standard sub-Eddington radiatively efficient accretion regime ($\eta_{\rm rad}=0.103$), and we further assumed that the spin is parallel to the angular momentum of the gas within the MBH kernel $l_{\rm gas,MBH}$.  While accretion is still modelled using the Bondi-Hoyle-Lyttleton prescription, the resulting MBH feedback was implemented in radiative and kinetic forms. 

Radiative feedback was injected as in \citet{lupi20b}, assuming a composite black body plus X-ray corona spectrum for the MBH, with a bolometric luminosity defined as $L_{\rm bol}=\eta_{\rm rad}\dot{M}_{\rm BH}c^2$, with $\eta_{\rm rad}$ depending on the accretion regime (see below). The fraction of $L_{\rm bol}$ associated with the X-ray corona was determined according to \citet{duras20} as
\begin{equation}
    f_{\rm 2-10\, keV}^{-1} \equiv\frac{L_{\rm bol}}{L_{\rm 2-10\, keV}} = 12.76\left[1+\left(\frac{\log(L_{\rm bol}/{\rm L_\odot})}{12.15}\right)^{18.78}\right]. 
\end{equation}
From this expression, we determined the soft X-ray fraction $f_{\rm 0.2-2\,\rm keV}$ assuming a power-law spectrum with a slope -1.7 for the corona \citep{regan19}, and we finally obtained the residual black-body component as 
         \begin{equation}
             f_{\rm BB} = 1- f_{\rm 0.2-2\,\rm keV}-f_{\rm 2-10\,\rm keV}.
         \end{equation} 
         
Kinetic feedback was instead implemented for the different regimes as\footnote{The final normalisations of the equations in this section correspond to the spin magnitude assumed in our simulation, $a=0.7$.}
\begin{itemize}
    \item $\lambda<2.5\times 10^{-3}$ \citep[ADAF regime,][]{yuan14}: In this regime, the disc is optically thin and geometrically thick due to inefficient cooling. Ions and electrons decouple, resulting in a two-temperature accretion flow where the radiative efficiency decays roughly as $\eta_{\rm rad}\propto \lambda^{2/3}$\citep{xie12}. In this regime, a jet is launched along $l_{\rm gas,MBH}$ and forms a cylinder with the base defined by the MBH kernel with an efficiency determined as \citep{blandfordznajek77,tchekhovskoy15}
    \begin{equation}
        \eta_{\rm jet}\approx 2.5\frac{a}{(1+\sqrt{1-a^2})^2}\left(\frac{\phi}{\phi_{\rm MAD}}\right)^2 \approx 0.417\Phi^2,
    \end{equation}
    where $\phi$ is the maximum magnetic flux in the disc, $\phi_{\rm MAD}$ is a critical value corresponding to a magnetically arrested disc \citep[MAD,][]{narayan03}, and $\Phi=\phi/\phi_{\rm MAD}$. 
    The mass loading of the jet $\beta_{\rm jet}\equiv \dot{M}_{\rm jet}/\dot{M}_{\rm BH}$ in our simulation was determined assuming energy conservation and a jet velocity $v_{\rm w}=0.1c=3\times 10^4\rm\, km\, s^{-1}$, which gives
     \begin{equation}
         \beta_{\rm jet} = \frac{\dot{M}_{\rm jet}}{\dot{M}_{\rm BH}} = 2\eta_{\rm jet}\left(\frac{c}{v_{\rm w}}\right)^2 \approx 83.4 \Phi^2 .
     \end{equation}
    
    \item $2.5\times 10^{-3}<\lambda<1$ (sub-Eddington radiatively efficient accretion): This regime occurs in typical AGN, where the disc is geometrically thin and optically thick, and it can be understood in terms of the Shakura and Sunyaev solution \citep{shakura73}. In this case, we assumed the MBH feedback to be in the form of radiation (again assuming a composite black-body plus corona spectrum) and bipolar line-driven winds whose mass loading is given by momentum conservation during the matter-radiation interaction \citep{choi12,anglesalcazar17FIRE}, that is,
    \begin{equation}
        \beta_{\rm w}= \frac{L_{\rm BH}}{\dot{M}_{\rm BH}v_{\rm w}c} = \eta_{\rm rad}\frac{c}{v_{\rm w}} \approx 1,
    \end{equation}
    where $L_{\rm BH} = \eta_{\rm rad}\dot{M}_{\rm BH}c^2$ is the accretion luminosity.
    The resulting energy coupling efficiency for the wind, which was launched assuming a semi-aperture of 45~$\deg$ \citep{sala21}, can be estimated in
    \begin{equation}
        \epsilon = \frac{\dot{M}_{\rm w}v_{\rm w}^2}{2L_{\rm BH}}=\frac{\dot{M}_{\rm w}v_{\rm w}^2}{2\eta_{\rm rad}\dot{M}_{\rm BH}c^2} = \frac{\beta v_{\rm w}^2}{2\eta_{\rm rad}c^2} = \frac{v_{\rm w}}{2c} = 0.05.
    \end{equation}
    \item $\lambda>1$ (super-Eddington accretion): In this regime, we assumed radiative plus kinetic MBH feedback, where the radiative component was determined according to the slim-disc solution, with the radiative efficiency obtained by \citet{lupi16},
    \begin{equation}
        \eta_{\rm rad}=\frac{A(a)}{16\lambda}\left[\frac{0.985}{1/\lambda+B(a)}+\frac{0.015}{1/\lambda+C(a)}\right],
    \end{equation}
    where $A(a)$, $B(a)$, and $C(a)$ are spin-dependent coefficients that in our case assumed values $1.915$, $0.795$, and $0.017$ respectively. Although radiation is expected to be trapped in the innermost regions of the slim disc, we neglected any change in the spectral shape of the radiation spectrum, leaving this exploration for a future study.
    The kinetic efficiency in this case was computed according to \citet{tchekhovskoy15} and \citet{sadowski16} for a magnetically driven jet, that is,
    \begin{equation}
        \eta_{\rm jet} = 1.3a^2\Phi^2 \approx 0.637\Phi^2, 
    \end{equation}
    corresponding to a mass loading $\beta_{\rm jet}\approx 127.4\Phi^2$.
\end{itemize}
A crucial parameter affecting the jet efficiency is the limiting magnetic field relative to the MAD limit. We considered two cases, one case at the MAD limit ($\Phi=1$) and the other case at half the limit ($\Phi=0.5$; HMAD hereon) \citep[see][for a discussion]{sadowski16}.
At every accretion event, we estimated the accretion rate on the MBH, ${\dot M}_{\rm BH},$ from its 96 nearest gas neighbours, compared it with the Eddington limit, and then determined $\eta_{\rm rad}$ and the value of $\beta$ for the corresponding regime. 

An important aspect that must be considered is that for very high accretion rates, the large mass-loading factor might yield a total gas mass affected by the MBH accretion/feedback process over a time step $\Delta t$ that exceeds the available mass in the MBH kernel. In these cases, we followed \citet{regan19} and assumed that the estimated accretion only occurred for a fraction of time,  
\begin{equation}
    f_{\rm acc} = \min\left\{1,\frac{M_{\rm ngbs}}{(1+\beta)\Delta M_{\rm BH}}\right\},
\end{equation}
where $\Delta M_{\rm BH}=\dot{M}_{\rm BH}\Delta t$ and $\beta$ is the mass-loading factor of the kinetic wind/jet.
At this point, we randomly flagged enough gas particles around the MBH to guarantee $\Delta M=(1+\beta) \dot{M}_{\rm BH}\Delta t$, where part of the mass is accreted and part is kicked away in a kinetic wind/jet. In the case of $M_{\rm dyn,MBH}>M_{\rm MBH}$, we simply selected particles to be kicked away in an outflow, in order to ensure mass conservation. In order to prevent the outflowing gas particles from propagating over large distances without interacting (over a single time step they might end up well outside the galaxy), we updated the time step $\Delta t$ of the kicked particles to $\Delta t' = \min\{\Delta t,C_{\rm CFL}\Delta x/v_{\rm w}\}$, where $\Delta x$ is the effective gas cell size \citep{hopkins15}, and $C_{\rm CFL}$ is the Courant factor.

\section{Results}
\label{sec:results}
\begin{figure*}
    \centering
    \includegraphics[width=\columnwidth]{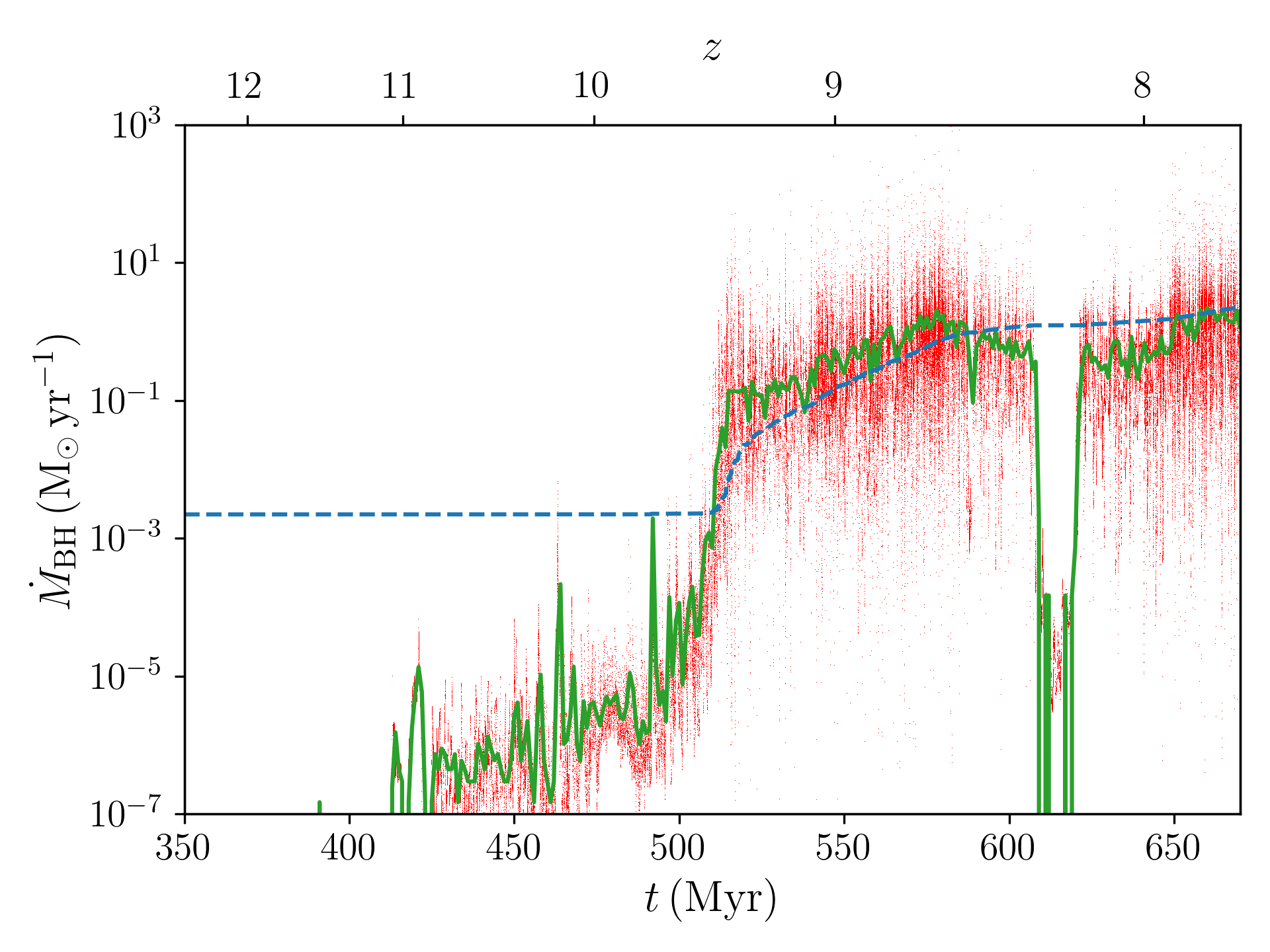}
    \includegraphics[width=\columnwidth]{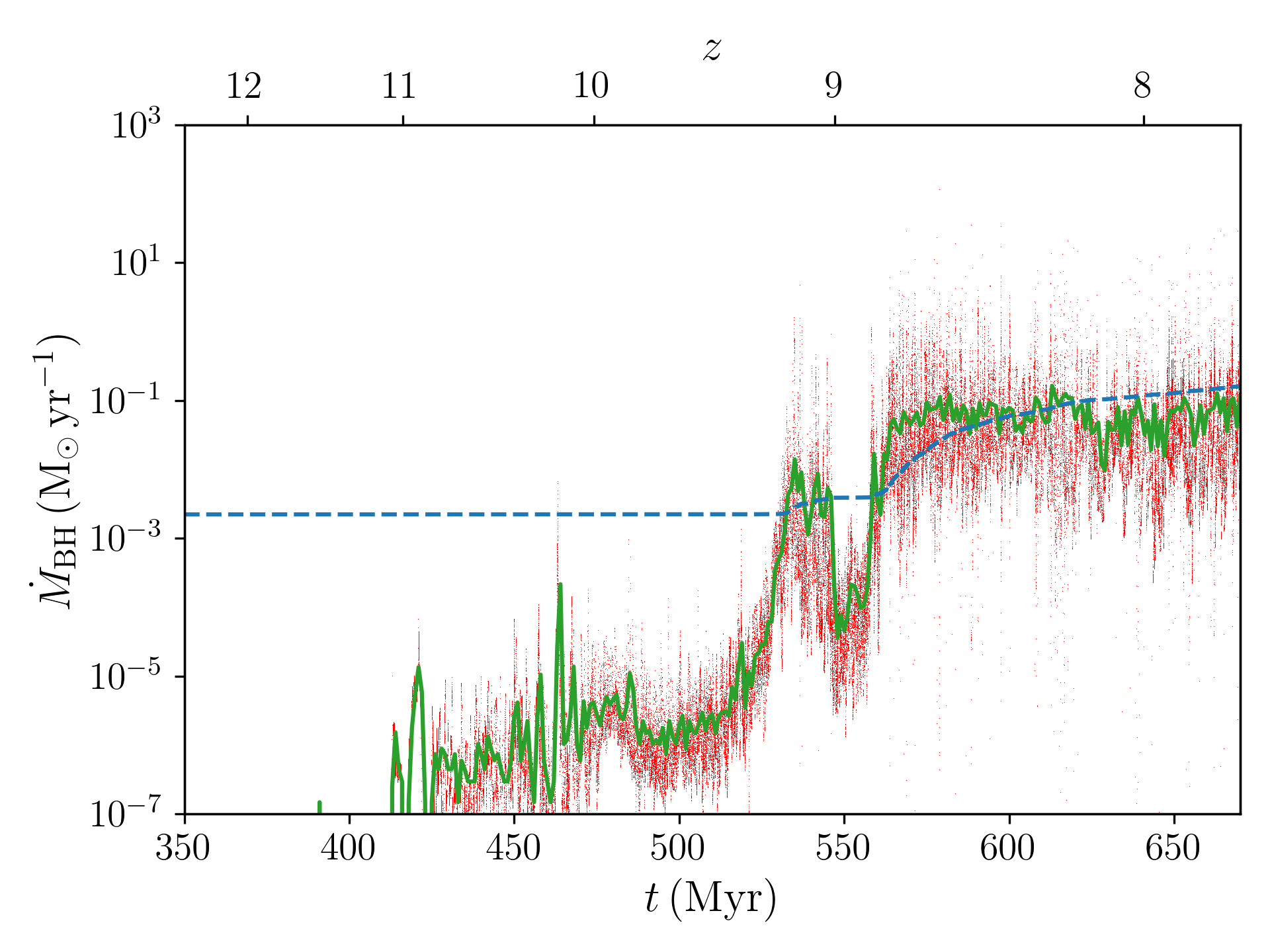}    
    \caption{Evolution of the MBH accretion rate in the HMAD (left panel) and MAD (right panel) simulations. The red dots correspond to the instantaneous accretion rate, i.e. at every time step of the simulations, whereas the solid green line corresponds to the accretion rate averaged over 1 Myr. The dashed blue line represents the instantaneous Eddington accretion rate at the MBH mass.}
    \label{fig:mdot}
\end{figure*}
With the model just described, we ran the same initial conditions of \citet{lupi19b} down to $z\sim 9.7$, when the MBH has already formed near the centre of its host, but the galaxy mass is still low enough for supernovae to stunt the MBH growth completely. At this point, we split the simulation into two equivalent runs, MAD and HMAD, which only differ by the magnetic flux parameter $\Phi$. The simulations were run down to $z\sim 7.5$ in order to follow the early growth of the MBH seeds in the galaxy host, whose evolution we discuss in this section.

In Fig.~\ref{fig:mdot}, we report the accretion rate on the MBH in the two cases, as directly obtained from the simulation (red dots) and averaged over a timescale of 1 Myr (solid green line). As a comparison, we also report as the dashed blue line the Eddington accretion rate at every step of the evolution. The instantaneous $\dot{M}_{\rm MBH}$ exhibits very large excursions, which are associated with the intermittency between large inflows and the MBH feedback self-regulation. 

The evolution can be divided into three main stages on average, however \citep[see also][]{lupi19b}. (i) In the early stages after MBH seeding ($t\lesssim 530$~Myr), supernova feedback strongly perturbs the gas in the galaxy, preventing the MBH (which is also offset from the galaxy centre) from efficiently accreting. (ii) When the MBH settles in the centre of the galaxy and the potential well becomes deep enough to sustain large gas inflows to the centre ($M_\star\simeq 10^{10}\msun$), $\dot{M}_{\rm MBH}$ rapidly grows, easily exceeding Eddington by a factor of a few (MAD) or a few tens (HMAD). At this stage, the HMAD simulation shows an almost unimpeded super-Eddington growth for about 60~Myr, followed by (iii) a decrease in the accretion rate to values between a fraction and 100\% Eddington, corresponding to the self-regulation stage. In both cases, there are sudden drops in the accretion rates at around $z= 8.3$ and $z=9$ for the HMAD and MAD cases, respectively. These rapid variations are associated with the potential misalignment of the gas in the galaxy nuclear region with respect to the galactic disc plane, which results in winds/jets that directly hit the disc instead of escaping perpendicular to it. In these phases, the gas is expelled from the galaxy centre and falls back on a timescale comparable with the free-fall time of the gas, that is, $5\,{\rm Myr}\lesssim t_{\rm ff}\lesssim \rm\, 50$~Myr for a typical gas density between 1 and 100~$\rm cm^{-3}$. 

\begin{figure*}
    \centering
    \includegraphics[width=\textwidth]{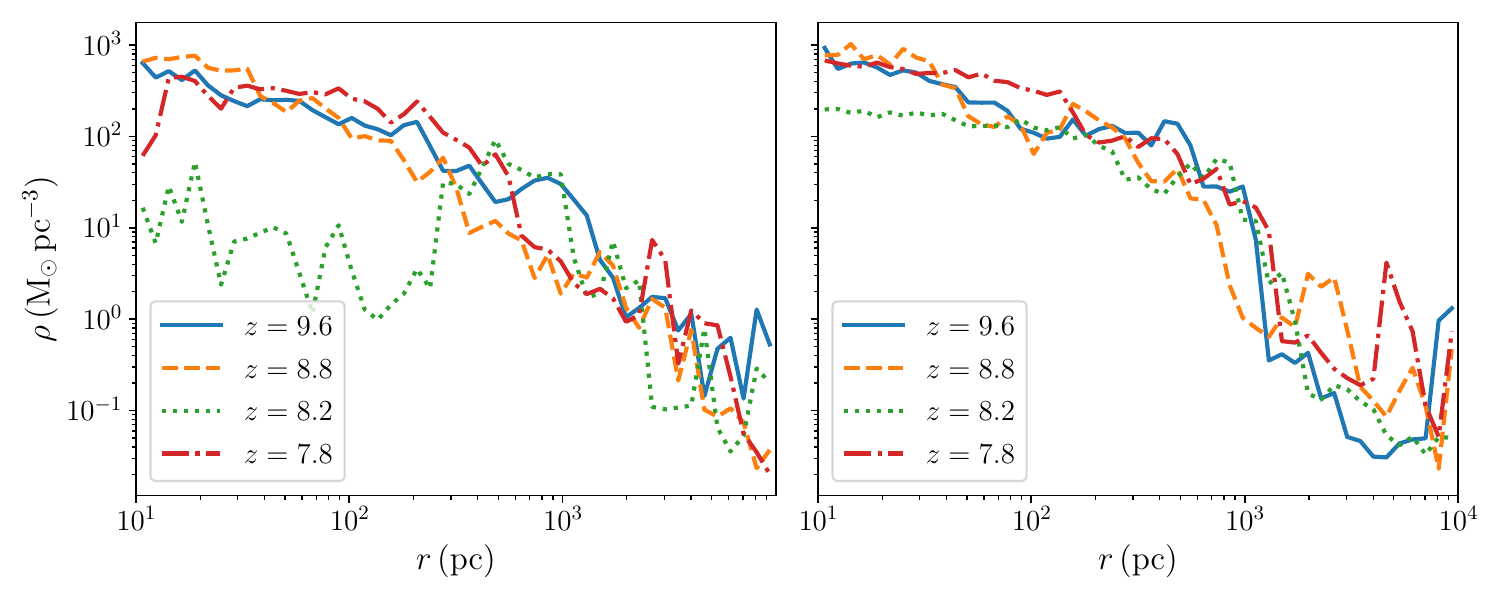}
    \caption{Gas density profiles for HMAD (left panel) and MAD (right panel) in spherical shells at different times: $z=9.6$ (sub-Eddington, solid blue line), $z=8.8$ (super-Eddington. dashed orange line), $z=8.2$ (not accreting in HMAD and mildly sub-Eddington in MAD, dotted green line), and $z=7.8$ (self-regulated, dot-dashed red line).}
    \label{fig:profile}
\end{figure*}

\begin{figure}
    \centering
    \includegraphics[width=\columnwidth]{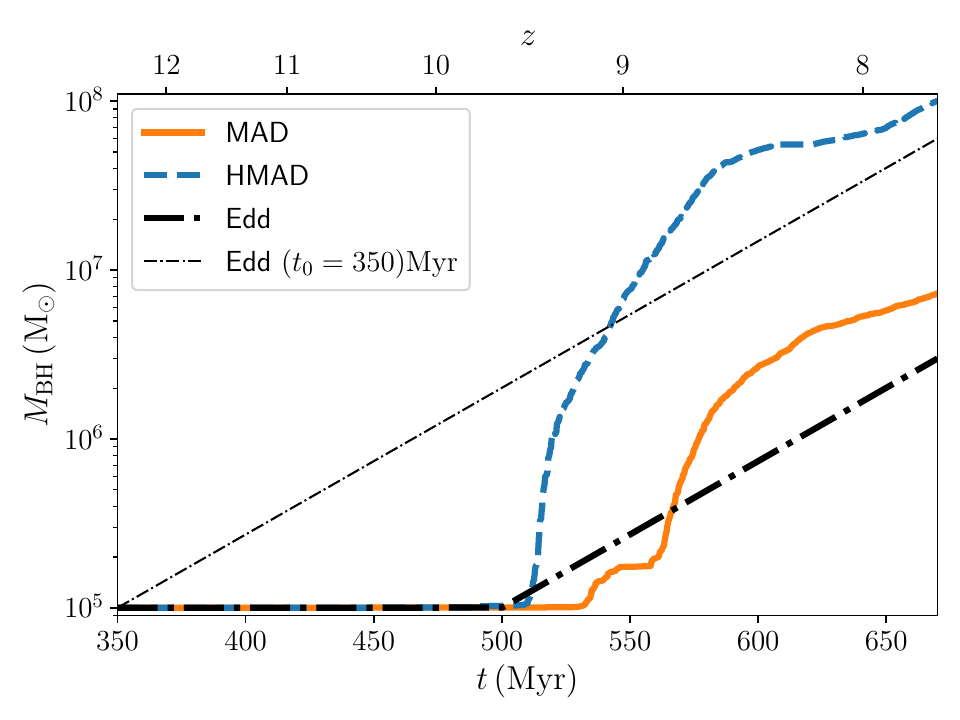}
    \caption{Evolution of the MBH mass in the MAD (shown as a solid orange line) and HMAD (shown as a dashed blue line) simulations. For comparison, we also show two dot-dashed black lines corresponding to the MBH growth for a constant Eddington accretion. The thick line starts at $t=500$~Myr (consistent with the HMAD case), and the thin line starts at $t=350$~Myr, immediately after seeding. }
    \label{fig:mass}
\end{figure}

The evolution just discussed reflects the mass growth reported in Fig.~\ref{fig:mass} for both simulations. The three stages just discussed can be noticed also in this figure. The time at which the MBH starts to grow is slightly different between the two simulations because of the small differences that appeared when the simulations were restarted from the parent simulation, which built up over time and changed the MBH history. 
Except for this and the plateaus in the growth history when the accretion rate drops, the growth exhibits a similar behaviour, that is, an initial rapid growth well above the Eddington limit, which lasts longer in the HMAD case due to the lower MBH feedback efficiency, followed by a self-regulation phase when the MBH settles at the Eddington limit or at a fraction of it. 

Despite the quite high efficiency of the jet feedback during super-Eddington phases, the HMAD MBH is able to grow by almost three orders of magnitude in less than 100~Myr, whereas the MAD MBH  stays one order of magnitude below throughout the same time interval. Another important aspect to keep in mind here is that when the MBH in the HMAD case starts to self-regulate its growth, the MBH mass is so high that super-Eddington phases begin to require extremely high inflow rates in the galaxy nucleus, which are increasingly less likely as the galaxy evolves and the gas fraction diminishes.
In the Eddington-limited case, an initially more massive MBH or an earlier start of accretion are needed to ensure a growth in mass that is comparable to the two cases we simulated here. In Fig.~\ref{fig:profile}, we report the  gas density profile from the two simulations (HMAD on the left and MAD on the right) computed in spherical shells at four different times: $z=9.6$ represents the initial growth phases, when the accretion rate in HMAD starts to rise towards the super-Eddington regime; $z=8.8$ is a super-Eddington phase in both simulations; $z=8.2$ represents a feedback-dominated phase, when the MBH in MAD slightly suppresses its growth, settling around the Eddington limit, whereas the MBH in HMAD launches a wind/jet through the disc that is able to carve a cavity of $\sim 300$~pc (as discussed above); and $z=7.8$, when both simulation settle around/below the Eddington limit, starting a self-regulated growth phase. Although the two simulations are formally independent, the gas distribution is in general very similar, but for the phases in which MBH feedback succeeds at expelling gas from the centre. This is consistent with most recent simulations that suggested that the central MBH does not directly affect the entire galaxy host, whose evolution is mainly determined (except for a few transient phases) by the cosmological environment.

\begin{figure}
    \centering
    \includegraphics[width=\columnwidth]{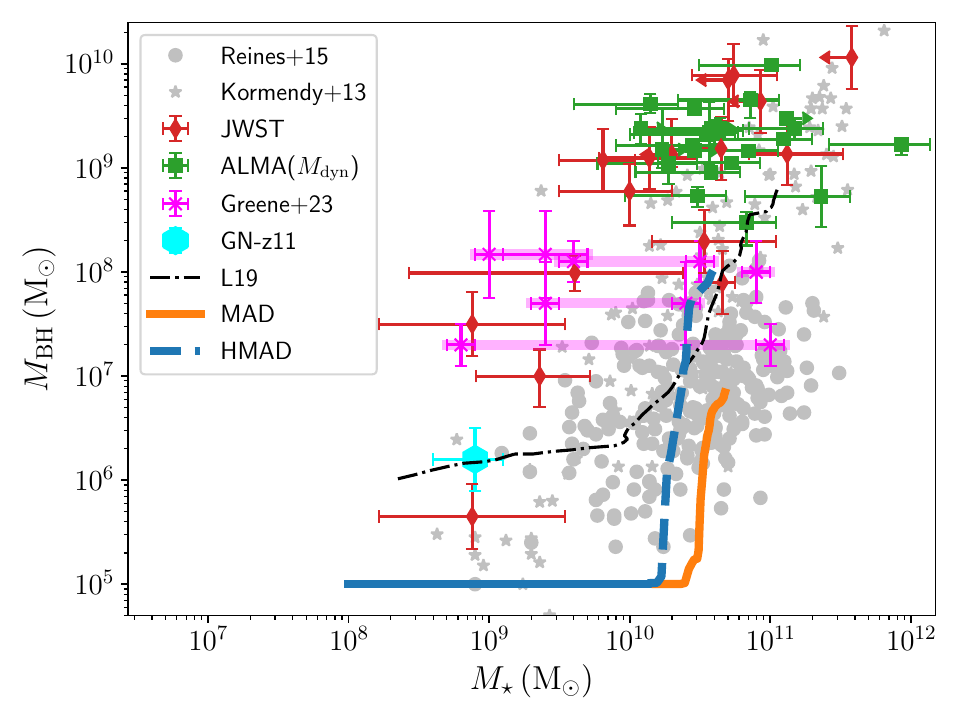}
    \caption{Evolution of the MBH mass in the MAD (shown as a solid orange line) and HMAD (shown as a dashed blue line) simulations relative to the galaxy stellar mass compared with the local AGN observations by \citet[][grey dots]{reines15}, local ellipticals by \citet[][grey stars]{kormendy13bh}, and high-redshift ($z\gtrsim 6$) observations by ALMA \citep[][green squares]{neeleman21} and JWST \citep[][red diamonds]{yue23,stone23a,stone24,harikane23,maiolino23b}. We also show the `little red dots' reported by \citet{greene24} as magenta crosses, with the stellar masses estimated with or without ALMA photometry, respectively, connected by magenta shaded areas \citep[see][for details]{labbe23}, and the $z\sim 10$ MBH GN-z11 \citep{maiolino24}, thought to be super-Eddington, as a cyan hexagon. For completeness, we also show the results of the simulation in L19 as a dot-dashed black line.}
    \label{fig:mbhmgal}
\end{figure}

Since this rapid growth may have important implications for the MBH-galaxy correlations, we report in Fig.~\ref{fig:mbhmgal} the evolution of the MBH together with the galaxy host stellar mass for the two cases, using the same colour scheme as in Fig.~\ref{fig:mass}. We also show for comparison the observations of low-redshift AGN by \citet{reines15} as grey dots, elliptical galaxies by \citet{kormendy13bh} as grey stars, and high-redshift (only at $z\gtrsim 6$) observations by ALMA as green squares \citep{neeleman21} and JWST as red diamonds \citep{yue23,stone23a,stone24,harikane23,maiolino23b}. We also report as magenta crosses the `little red dots' observed by \citet{greene24}, with the stellar masses estimated according to JWST photometry alone and including ALMA photometry by \citet{labbe23}, where the magenta shaded area connects the two mass estimates for each source.\footnote{The emission in this case is dominated by the AGN and dust, hence, the stellar mass estimates are only indicative, but likely lower than 10$^{10}\msun$ because the little red dots are selected to be point sources.} Finally, we also report as a cyan hexagon GN-z11 \citep{maiolino24}, the highest-redshift MBH to date, which is thought to be accreting at super-Eddington rates, and the results by L19 as a dot-dashed black line. The overlap between GN-z11 and the L19 simulation is simply due to the higher seeding mass ($10^6\rm\, \msun$) in L19 and not to the physical evolution in the simulation. In both cases, the MBH starts at the lower end of the distribution and grows rapidly, but only the HMAD case is able to reach the upper limit of the distribution because of a less effective MBH feedback. Then, when the MBH starts to self-regulate, even the growth of the HMAD MBH slows down, moving toward the region occupied by giant ellipticals (see L19 for a discussion).

Finally, we note that the super-Eddington phase was able to start only after the weakening of the impact of supernova feedback on the galaxy, in agreement with other simulations \citep{dubois14,anglesalcazar17FIRE,lupi19b}. This prevented the simulated MBHs from reaching the region occupied by high-redshift over-massive MBHs, especially those observed in low-mass galaxies \citep[e.g.][]{maiolino23b}. This might be a potential issue in explaining the overmassive MBHs, unless an even weaker super-Eddington feedback or more favourable inflow conditions occur around these MBHs.

\begin{figure*}
    \centering
    \includegraphics[width=\columnwidth]{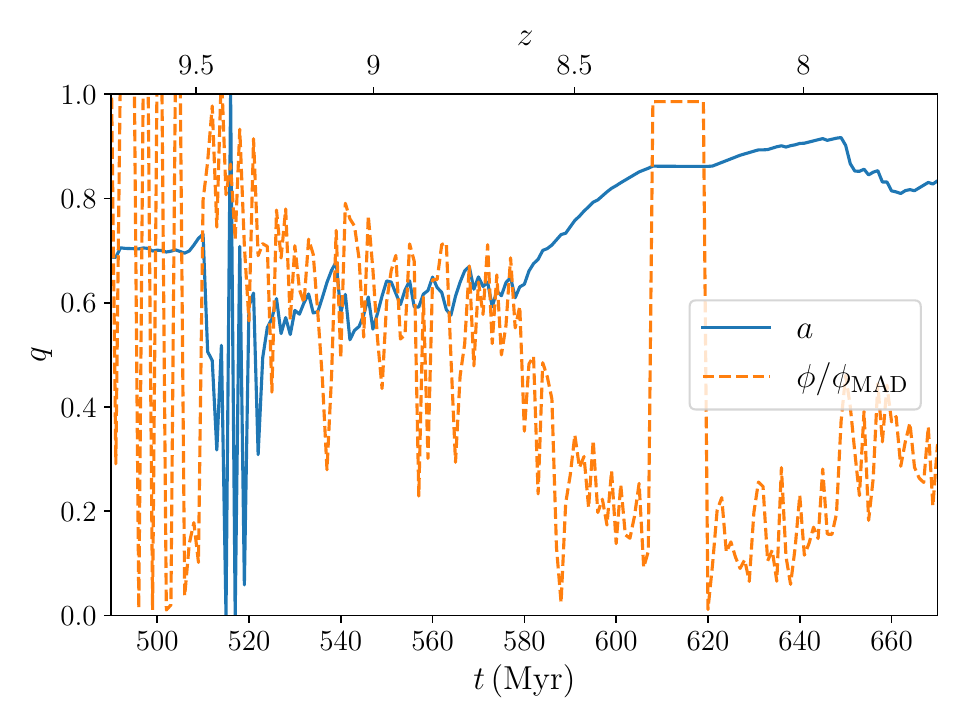}
    \includegraphics[width=\columnwidth]{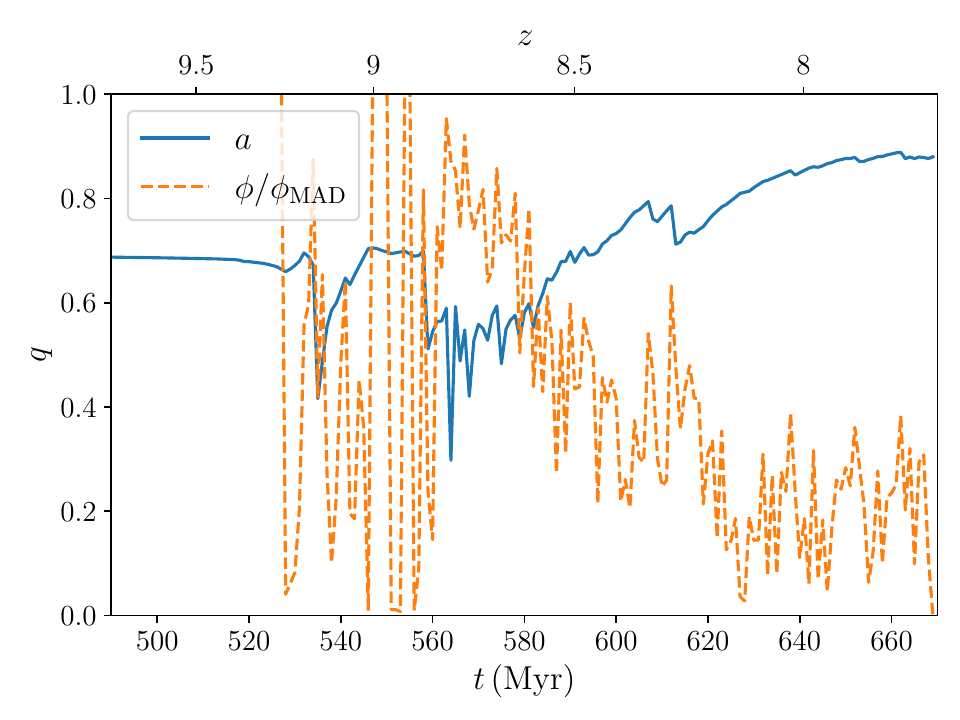}    
    \caption{Evolution of the MBH spin (solid blue line) and MAD-ness (dashed orange line) in the HMAD (left panel) and MAD (right panel) simulations, as modelled in \citet{ricarte23}.}
    \label{fig:spinup}
\end{figure*}
\section{Discussion and conclusions}
\label{sec:conclusions}

We have performed high-resolution cosmological zoom-in simulations of a massive halo at high redshift that is expected to be a quasar host. The simulations include for the first time in a cosmological simulation a sub-resolution model for the three main accretion regimes, from  ADAF up to  super-Eddington. The model accounts for MBH accretion-powered feedback in the form of radiation and kinetic winds/jets, with an efficiency that is determined consistently for the different cases. The results we found here will be complemented by a more thorough analysis of the MBH-galaxy interplay and the properties of the MBH outflows in a forthcoming paper (Quadri et al. in prep.). They show that super-Eddington phases might be sustained over timescales of a few tens of millions of years in massive systems where large gas inflows are frequent, systems that are quite different from those investigated in previous works. However, some important caveats are listed below. 
\begin{itemize}
\item The resolution of the simulation is not high enough to properly resolve the Bondi-Hoyle-Lyttleton radius of the MBH. This might affect the estimated accretion rate.  Nonetheless, a higher accretion rate is reflected in a stronger feedback, which has the potential of clearing out the central region of the galaxy (as sometimes occurred in our simulations). This would hinder accretion for a few tens of millions of years.
\item The launch direction of the kinetic feedback is instantaneously aligned with the gas within the MBH kernel, without taking into account the actual alignment of an unresolved accretion disc. This means that if the gas in the galaxy has settled into a disc-like configuration, the kinetic feedback escapes most of the time without interacting significantly with the galaxy, which means that accretion is not suppressed for a long time. In a more realistic case, the alignment of the accretion disc might occur on much longer timescales and might in this way affect the impact of the feedback.  

\item 
The results of this simulation provide a more optimistic outlook on the role of super-Eddington accretion in growing high-redshift black holes compared to previous numerical investigations in a galactic/cosmological context \citep{johnson11,regan19,massonneau23}. In particular, even though \citet{johnson11} studied the early growth of MBH seeds in highly overdense regions, finding that radiation is able to regulate MBH accretion, their exploration was limited to the very first stages, that is, a pristine atomic cooling halo that grew little mass over the 1~Myr timescale they considered. In our case, the environment in which the MBH is embedded is completely different, as it is a metal-enriched massive galaxy with massive inflows. In metal-rich conditions and at high densities, gas cooling becomes very efficient and thus suppresses the effect of accretion-powered photoheating \citep[in agreement with the ineffective super-Eddington growth found by][in similar physical conditions]{regan19}. To assess the reasons for the difference with \citet{regan19} and \citet{massonneau23}, we  measured the gas density and temperature around the MBH, finding that super-Eddington bursts were followed by an increase in the gas temperature up to $10^6<T/{\rm K}<10^9$ (above the peak of the cooling curve) due to the jet shocking with the gas. This is consistent with the results in \citet{regan19} and \citet{massonneau23}. Despite this strong heating, the density around the MBH did not change dramatically and always remained in the range $n_{\rm H}\simeq 10^{4-5}\rm\, cm^{-3}$ on average. This value is significantly higher than that in \citealt{massonneau23}. This suggests that the combination of a deep potential well, the pressure of the inflowing gas from larger scales, and a short free-fall time (see \citealt{regan19} for a discussion) prevented the gas from escaping from the galaxy nucleus. As a consequence, after each jet event, the accretion rate suppression we observed was short-lived and was followed by a rapid rise as soon as the gas cooled down. However, the limited resolution of our simulation might still underestimate the effect of the jet feedback by not properly resolving the early expansion of the shocked gas, and further investigations are needed.
\item The magnitude of the spin in the simulations was kept fixed. This has profound implications for the evolution. First, since the jet launched during super-Eddington accretion phases extracts rotational energy from the MBH \citep{blandfordznajek77}, the spin should decrease rapidly, thus overcoming the spin-up due to coherent accretion. This would likely result in a lower spin during the first accretion phases and hence, in a lower kinetic efficiency
for the jet, even in the MAD case, which would enable the MBH to grow almost unimpededly until the MBH spin and mass become high enough for feedback to start playing a role \citep{massonneau23spin}. 
In order to test this idea, we computed the spin evolution semi-analytically by means of the model in \citet{ricarte23}, employing the accretion rates and the MBH masses from our simulations. In particular, we employed the 1~Myr average quantities instead of the full data sample in order to limit the effect of strong fluctuations on our conclusions. The results are reported in Fig.~\ref{fig:spinup}, where we show the evolution of the MBH spin and of the disc MAD-ness as defined in \citet{ricarte23} for $f_{\rm Edd}>0.03$ (at lower Eddington ratios, a MAD disc is assumed),
\begin{equation}
    \frac{\phi}{\phi_{\rm MAD}} =
     \frac{(f_{\rm Edd}/f_{\rm c})^\alpha}{1+(f_{\rm Edd}/f_{\rm c})^\alpha}, 
\end{equation}
where $f_{\rm c}=1.88$ and $\alpha=1.29$. 

We clearly observe that during the super-Eddington phases, the MBH spin decreases to values around 0.5--0.6, which correspond to jet efficiencies that are lower by 1.5--2 times than those employed in the simulations. Only during the later phases, when the accretion rate settles at sub-Eddington values, the spin is able to grow beyond the initial value. During the evolution, the MAD-ness of the disc also exhibits strong variations, with average values during the super-Eddington phases of about 0.6 (HMAD) and 0.8 (MAD), which is in between the values assumed in our simulations. Combining these effects, we obtain a net efficiency of the super-Eddington jets of $
\eta_{\rm jet}\propto a^2(\phi/\phi_{\rm MAD})^2$ , which is more in line with our HMAD simulation than with the MAD case.
\end{itemize}
To conclude, we have shown that despite the unfavourable conditions associated with assuming a relatively high spin and a consequently strong jet feedback, MBHs in gas-rich environments at high redshift can support long-lasting super-Eddington accretion phases and grow rapidly in mass within their host galaxies. 
Moreover, because of this rapid growth, these MBHs can easily move above the local correlations before they start to regulate themselves. This might explain the formation of overmassive systems.

\begin{acknowledgements}
AL and MC acknowledge support by the PRIN MUR "2022935STW". JR acknowledges support from the Royal Society and Science Foundation Ireland under grant number URF\textbackslash R1\textbackslash 191132. JR also acknowledges support from the Irish Research Council Laureate programme under grant number IRCLA/2022/1165.
\end{acknowledgements}

\bibliographystyle{aa}
\bibliography{./Biblio}

\begin{thebibliography}{86}
\expandafter\ifx\csname natexlab\endcsname\relax\def\natexlab#1{#1}\fi

\bibitem[{{Abramowicz} {et~al.}(1988){Abramowicz}, {Czerny}, {Lasota}, \&
  {Szuszkiewicz}}]{abramowicz88}
{Abramowicz}, M.~A., {Czerny}, B., {Lasota}, J.~P., \& {Szuszkiewicz}, E. 1988,
  \apj, 332, 646

\bibitem[{{Angl{\'e}s-Alc{\'a}zar} {et~al.}(2017){Angl{\'e}s-Alc{\'a}zar},
  {Faucher-Gigu{\`e}re}, {Quataert}, {Hopkins}, {Feldmann}, {Torrey}, {Wetzel},
  \& {Kere{\v s}}}]{anglesalcazar17FIRE}
{Angl{\'e}s-Alc{\'a}zar}, D., {Faucher-Gigu{\`e}re}, C.-A., {Quataert}, E.,
  {et~al.} 2017, \mnras, 472, L109

\bibitem[{{Ba{\~n}ados} {et~al.}(2018){Ba{\~n}ados}, {Venemans},
  {Mazzucchelli}, {Farina}, {Walter}, {Wang}, {Decarli}, {Stern}, {Fan},
  {Davies}, {Hennawi}, {Simcoe}, {Turner}, {Rix}, {Yang}, {Kelson}, {Rudie}, \&
  {Winters}}]{banados18}
{Ba{\~n}ados}, E., {Venemans}, B.~P., {Mazzucchelli}, C., {et~al.} 2018, \nat,
  553, 473

\bibitem[{{Bachetti} {et~al.}(2014){Bachetti}, {Harrison}, {Walton},
  {Grefenstette}, {Chakrabarty}, {F{\"u}rst}, {Barret}, {Beloborodov}, {Boggs},
  {Christensen}, {Craig}, {Fabian}, {Hailey}, {Hornschemeier}, {Kaspi},
  {Kulkarni}, {Maccarone}, {Miller}, {Rana}, {Stern}, {Tendulkar}, {Tomsick},
  {Webb}, \& {Zhang}}]{bachetti14}
{Bachetti}, M., {Harrison}, F.~A., {Walton}, D.~J., {et~al.} 2014, \nat, 514,
  202

\bibitem[{{Barai} {et~al.}(2018){Barai}, {Gallerani}, {Pallottini}, {Ferrara},
  {Marconi}, {Cicone}, {Maiolino}, \& {Carniani}}]{barai18}
{Barai}, P., {Gallerani}, S., {Pallottini}, A., {et~al.} 2018, \mnras, 473,
  4003

\bibitem[{{Bhowmick} {et~al.}(2022){Bhowmick}, {Blecha}, {Ni}, {Di Matteo},
  {Torrey}, {Kelley}, {Vogelsberger}, {Weinberger}, \&
  {Hernquist}}]{bhowmick22}
{Bhowmick}, A.~K., {Blecha}, L., {Ni}, Y., {et~al.} 2022, \mnras, 516, 138

\bibitem[{{Blandford} \& {Znajek}(1977)}]{blandfordznajek77}
{Blandford}, R.~D. \& {Znajek}, R.~L. 1977, \mnras, 179, 433

\bibitem[{{Choi} {et~al.}(2012){Choi}, {Ostriker}, {Naab}, \&
  {Johansson}}]{choi12}
{Choi}, E., {Ostriker}, J.~P., {Naab}, T., \& {Johansson}, P.~H. 2012, \apj,
  754, 125

\bibitem[{{Davies} {et~al.}(2019){Davies}, {Hennawi}, \& {Eilers}}]{davies19}
{Davies}, F.~B., {Hennawi}, J.~F., \& {Eilers}, A.-C. 2019, \apjl, 884, L19

\bibitem[{{Di Matteo} {et~al.}(2017){Di Matteo}, {Croft}, {Feng}, {Waters}, \&
  {Wilkins}}]{dimatteo17}
{Di Matteo}, T., {Croft}, R.~A.~C., {Feng}, Y., {Waters}, D., \& {Wilkins}, S.
  2017, \mnras, 467, 4243

\bibitem[{{Di Matteo} {et~al.}(2005){Di Matteo}, {Springel}, \&
  {Hernquist}}]{dimatteo05}
{Di Matteo}, T., {Springel}, V., \& {Hernquist}, L. 2005, \nat, 433, 604

\bibitem[{{Dubois} {et~al.}(2013){Dubois}, {Gavazzi}, {Peirani}, \&
  {Silk}}]{dubois13}
{Dubois}, Y., {Gavazzi}, R., {Peirani}, S., \& {Silk}, J. 2013, \mnras, 433,
  3297

\bibitem[{{Dubois} {et~al.}(2014){Dubois}, {Pichon}, {Welker}, {Le Borgne},
  {Devriendt}, {Laigle}, {Codis}, {Pogosyan}, {Arnouts}, {Benabed}, {Bertin},
  {Blaizot}, {Bouchet}, {Cardoso}, {Colombi}, {de Lapparent}, {Desjacques},
  {Gavazzi}, {Kassin}, {Kimm}, {McCracken}, {Milliard}, {Peirani}, {Prunet},
  {Rouberol}, {Silk}, {Slyz}, {Sousbie}, {Teyssier}, {Tresse}, {Treyer},
  {Vibert}, \& {Volonteri}}]{dubois14}
{Dubois}, Y., {Pichon}, C., {Welker}, C., {et~al.} 2014, \mnras, 444, 1453

\bibitem[{{Duras} {et~al.}(2020){Duras}, {Bongiorno}, {Ricci}, {Piconcelli},
  {Shankar}, {Lusso}, {Bianchi}, {Fiore}, {Maiolino}, {Marconi}, {Onori},
  {Sani}, {Schneider}, {Vignali}, \& {La Franca}}]{duras20}
{Duras}, F., {Bongiorno}, A., {Ricci}, F., {et~al.} 2020, \aap, 636, A73

\bibitem[{{Escala} {et~al.}(2004){Escala}, {Larson}, {Coppi}, \&
  {Mardones}}]{escala04}
{Escala}, A., {Larson}, R.~B., {Coppi}, P.~S., \& {Mardones}, D. 2004, \apj,
  607, 765

\bibitem[{{Fan} {et~al.}(2023){Fan}, {Ba{\~n}ados}, \& {Simcoe}}]{fan23}
{Fan}, X., {Ba{\~n}ados}, E., \& {Simcoe}, R.~A. 2023, \araa, 61, 373

\bibitem[{{Fan} {et~al.}(2006){Fan}, {Strauss}, {Richards}, {Hennawi},
  {Becker}, {White}, {Diamond-Stanic}, {Donley}, {Jiang}, {Kim}, {Vestergaard},
  {Young}, {Gunn}, {Lupton}, {Knapp}, {Schneider}, {Brandt}, {Bahcall},
  {Barentine}, {Brinkmann}, {Brewington}, {Fukugita}, {Harvanek}, {Kleinman},
  {Krzesinski}, {Long}, {Neilsen}, {Nitta}, {Snedden}, \& {Voges}}]{fan06}
{Fan}, X., {Strauss}, M.~A., {Richards}, G.~T., {et~al.} 2006, \aj, 131, 1203

\bibitem[{{Fiacconi} {et~al.}(2017){Fiacconi}, {Mayer}, {Madau}, {Lupi},
  {Dotti}, \& {Haardt}}]{fiacconi17}
{Fiacconi}, D., {Mayer}, L., {Madau}, P., {et~al.} 2017, \mnras, 467, 4080

\bibitem[{{Greene} {et~al.}(2024){Greene}, {Labbe}, {Goulding}, {Furtak},
  {Chemerynska}, {Kokorev}, {Dayal}, {Volonteri}, {Williams}, {Wang}, {Setton},
  {Burgasser}, {Bezanson}, {Atek}, {Brammer}, {Cutler}, {Feldmann}, {Fujimoto},
  {Glazebrook}, {de Graaff}, {Khullar}, {Leja}, {Marchesini}, {Maseda},
  {Matthee}, {Miller}, {Naidu}, {Nanayakkara}, {Oesch}, {Pan}, {Papovich},
  {Price}, {van Dokkum}, {Weaver}, {Whitaker}, \& {Zitrin}}]{greene24}
{Greene}, J.~E., {Labbe}, I., {Goulding}, A.~D., {et~al.} 2024, \apj, 964, 39

\bibitem[{{Hahn} \& {Abel}(2013)}]{hahn13}
{Hahn}, O. \& {Abel}, T. 2013, {MUSIC: MUlti-Scale Initial Conditions},
  Astrophysics Source Code Library

\bibitem[{{Harikane} {et~al.}(2023){Harikane}, {Zhang}, {Nakajima}, {Ouchi},
  {Isobe}, {Ono}, {Hatano}, {Xu}, \& {Umeda}}]{harikane23}
{Harikane}, Y., {Zhang}, Y., {Nakajima}, K., {et~al.} 2023, \apj, 959, 39

\bibitem[{{Hopkins}(2015)}]{hopkins15}
{Hopkins}, P.~F. 2015, \mnras, 450, 53

\bibitem[{{Hopkins} {et~al.}(2013){Hopkins}, {Cox}, {Hernquist}, {Narayanan},
  {Hayward}, \& {Murray}}]{hopkins13sf}
{Hopkins}, P.~F., {Cox}, T.~J., {Hernquist}, L., {et~al.} 2013, \mnras, 430,
  1901

\bibitem[{{Inayoshi} {et~al.}(2020){Inayoshi}, {Visbal}, \&
  {Haiman}}]{Inayoshi20}
{Inayoshi}, K., {Visbal}, E., \& {Haiman}, Z. 2020, \araa, 58, 27

\bibitem[{{Jiang} {et~al.}(2019){Jiang}, {Blaes}, {Stone}, \&
  {Davis}}]{jiang19}
{Jiang}, Y.-F., {Blaes}, O., {Stone}, J.~M., \& {Davis}, S.~W. 2019, \apj, 885,
  144

\bibitem[{{Johnson} {et~al.}(2011){Johnson}, {Khochfar}, {Greif}, \&
  {Durier}}]{johnson11}
{Johnson}, J.~L., {Khochfar}, S., {Greif}, T.~H., \& {Durier}, F. 2011, \mnras,
  410, 919

\bibitem[{{Kormendy} \& {Ho}(2013)}]{kormendy13bh}
{Kormendy}, J. \& {Ho}, L.~C. 2013, \araa, 51, 511

\bibitem[{{Labbe} {et~al.}(2023){Labbe}, {Greene}, {Bezanson}, {Fujimoto},
  {Furtak}, {Goulding}, {Matthee}, {Naidu}, {Oesch}, {Atek}, {Brammer},
  {Chemerynska}, {Coe}, {Cutler}, {Dayal}, {Feldmann}, {Franx}, {Glazebrook},
  {Leja}, {Marchesini}, {Maseda}, {Nanayakkara}, {Nelson}, {Pan}, {Papovich},
  {Price}, {Suess}, {Wang}, {Whitaker}, {Williams}, \& {Zitrin}}]{labbe23}
{Labbe}, I., {Greene}, J.~E., {Bezanson}, R., {et~al.} 2023, arXiv e-prints,
  arXiv:2306.07320

\bibitem[{{Latif} {et~al.}(2015){Latif}, {Bovino}, {Grassi}, {Schleicher}, \&
  {Spaans}}]{latif15}
{Latif}, M.~A., {Bovino}, S., {Grassi}, T., {Schleicher}, D.~R.~G., \&
  {Spaans}, M. 2015, \mnras, 446, 3163

\bibitem[{{Latif} {et~al.}(2022){Latif}, {Whalen}, {Khochfar}, {Herrington}, \&
  {Woods}}]{latif22}
{Latif}, M.~A., {Whalen}, D.~J., {Khochfar}, S., {Herrington}, N.~P., \&
  {Woods}, T.~E. 2022, \nat, 607, 48

\bibitem[{{Lin} {et~al.}(2017){Lin}, {Guillochon}, {Komossa}, {Ramirez-Ruiz},
  {Irwin}, {Maksym}, {Grupe}, {Godet}, {Webb}, {Barret}, {Zauderer}, {Duc},
  {Carrasco}, \& {Gwyn}}]{lin17}
{Lin}, D., {Guillochon}, J., {Komossa}, S., {et~al.} 2017, Nature Astronomy, 1,
  0033

\bibitem[{{Lupi} \& {Bovino}(2020)}]{lupi20}
{Lupi}, A. \& {Bovino}, S. 2020, \mnras, 492, 2818

\bibitem[{{Lupi} {et~al.}(2016){Lupi}, {Haardt}, {Dotti}, {Fiacconi}, {Mayer},
  \& {Madau}}]{lupi16}
{Lupi}, A., {Haardt}, F., {Dotti}, M., {et~al.} 2016, \mnras, 456, 2993

\bibitem[{{Lupi} {et~al.}(2021){Lupi}, {Haiman}, \& {Volonteri}}]{lupi21}
{Lupi}, A., {Haiman}, Z., \& {Volonteri}, M. 2021, \mnras, 503, 5046

\bibitem[{{Lupi} {et~al.}(2020){Lupi}, {Pallottini}, {Ferrara}, {Bovino},
  {Carniani}, \& {Vallini}}]{lupi20b}
{Lupi}, A., {Pallottini}, A., {Ferrara}, A., {et~al.} 2020, \mnras, 496, 5160

\bibitem[{{Lupi} {et~al.}(2022){Lupi}, {Volonteri}, {Decarli}, {Bovino}, \&
  {Silk}}]{lupi22}
{Lupi}, A., {Volonteri}, M., {Decarli}, R., {Bovino}, S., \& {Silk}, J. 2022,
  \mnras, 510, 5760

\bibitem[{{Lupi} {et~al.}(2019){Lupi}, {Volonteri}, {Decarli}, {Bovino},
  {Silk}, \& {Bergeron}}]{lupi19b}
{Lupi}, A., {Volonteri}, M., {Decarli}, R., {et~al.} 2019, \mnras, 488, 4004

\bibitem[{{Ma} {et~al.}(2021){Ma}, {Hopkins}, {Ma}, {Angl{\'e}s-Alc{\'a}zar},
  {Faucher-Gigu{\`e}re}, \& {Kelley}}]{ma21}
{Ma}, L., {Hopkins}, P.~F., {Ma}, X., {et~al.} 2021, \mnras, 508, 1973

\bibitem[{{Madau} {et~al.}(2014){Madau}, {Haardt}, \& {Dotti}}]{madau14}
{Madau}, P., {Haardt}, F., \& {Dotti}, M. 2014, \apjl, 784, L38

\bibitem[{{Maiolino} {et~al.}(2023){Maiolino}, {Scholtz}, {Curtis-Lake},
  {Carniani}, {Baker}, {de Graaff}, {Tacchella}, {{\"U}bler}, {D'Eugenio},
  {Witstok}, {Curti}, {Arribas}, {Bunker}, {Charlot}, {Chevallard},
  {Eisenstein}, {Egami}, {Ji}, {Jones}, {Lyu}, {Rawle}, {Robertson},
  {Rujopakarn}, {Perna}, {Sun}, {Venturi}, {Williams}, \&
  {Willott}}]{maiolino23b}
{Maiolino}, R., {Scholtz}, J., {Curtis-Lake}, E., {et~al.} 2023, arXiv
  e-prints, arXiv:2308.01230

\bibitem[{{Maiolino} {et~al.}(2024){Maiolino}, {Scholtz}, {Witstok},
  {Carniani}, {D'Eugenio}, {de Graaff}, {{\"U}bler}, {Tacchella},
  {Curtis-Lake}, {Arribas}, {Bunker}, {Charlot}, {Chevallard}, {Curti},
  {Looser}, {Maseda}, {Rawle}, {Rodr{\'\i}guez del Pino}, {Willott}, {Egami},
  {Eisenstein}, {Hainline}, {Robertson}, {Williams}, {Willmer}, {Baker},
  {Boyett}, {DeCoursey}, {Fabian}, {Helton}, {Ji}, {Jones}, {Kumari},
  {Laporte}, {Nelson}, {Perna}, {Sandles}, {Shivaei}, \& {Sun}}]{maiolino24}
{Maiolino}, R., {Scholtz}, J., {Witstok}, J., {et~al.} 2024, \nat, 627, 59

\bibitem[{{Marconi} {et~al.}(2004){Marconi}, {Risaliti}, {Gilli}, {Hunt},
  {Maiolino}, \& {Salvati}}]{marconi2004}
{Marconi}, A., {Risaliti}, G., {Gilli}, R., {et~al.} 2004, \mnras, 351, 169

\bibitem[{{Martizzi} {et~al.}(2015){Martizzi}, {Faucher-Gigu{\`e}re}, \&
  {Quataert}}]{martizzi15}
{Martizzi}, D., {Faucher-Gigu{\`e}re}, C.-A., \& {Quataert}, E. 2015, \mnras,
  450, 504

\bibitem[{{Massonneau} {et~al.}(2023{\natexlab{a}}){Massonneau}, {Dubois},
  {Volonteri}, \& {Beckmann}}]{massonneau23spin}
{Massonneau}, W., {Dubois}, Y., {Volonteri}, M., \& {Beckmann}, R.~S.
  2023{\natexlab{a}}, \aap, 669, A143

\bibitem[{{Massonneau} {et~al.}(2023{\natexlab{b}}){Massonneau}, {Volonteri},
  {Dubois}, \& {Beckmann}}]{massonneau23}
{Massonneau}, W., {Volonteri}, M., {Dubois}, Y., \& {Beckmann}, R.~S.
  2023{\natexlab{b}}, \aap, 670, A180

\bibitem[{{Mignoli} {et~al.}(2020){Mignoli}, {Gilli}, {Decarli}, {Vanzella},
  {Balmaverde}, {Cappelluti}, {Cassar{\`a}}, {Comastri}, {Cusano}, {Iwasawa},
  {Marchesi}, {Prandoni}, {Vignali}, {Vito}, {Zamorani}, {Chiaberge}, \&
  {Norman}}]{mignoli20}
{Mignoli}, M., {Gilli}, R., {Decarli}, R., {et~al.} 2020, \aap, 642, L1

\bibitem[{{Mortlock} {et~al.}(2011){Mortlock}, {Warren}, {Venemans}, {Patel},
  {Hewett}, {McMahon}, {Simpson}, {Theuns}, {Gonz{\'a}les-Solares}, {Adamson},
  {Dye}, {Hambly}, {Hirst}, {Irwin}, {Kuiper}, {Lawrence}, \&
  {R{\"o}ttgering}}]{mortlock11}
{Mortlock}, D.~J., {Warren}, S.~J., {Venemans}, B.~P., {et~al.} 2011, \nat,
  474, 616

\bibitem[{{Narayan} {et~al.}(2003){Narayan}, {Igumenshchev}, \&
  {Abramowicz}}]{narayan03}
{Narayan}, R., {Igumenshchev}, I.~V., \& {Abramowicz}, M.~A. 2003, \pasj, 55,
  L69

\bibitem[{{Neeleman} {et~al.}(2021){Neeleman}, {Novak}, {Venemans}, {Walter},
  {Decarli}, {Kaasinen}, {Schindler}, {Ba{\~n}ados}, {Carilli}, {Drake}, {Fan},
  \& {Rix}}]{neeleman21}
{Neeleman}, M., {Novak}, M., {Venemans}, B.~P., {et~al.} 2021, \apj, 911, 141

\bibitem[{{Neumayer} {et~al.}(2020){Neumayer}, {Seth}, \&
  {B{\"o}ker}}]{neumayer20}
{Neumayer}, N., {Seth}, A., \& {B{\"o}ker}, T. 2020, \aapr, 28, 4

\bibitem[{{Ni} {et~al.}(2022){Ni}, {Di Matteo}, {Bird}, {Croft}, {Feng},
  {Chen}, {Tremmel}, {DeGraf}, \& {Li}}]{ni22}
{Ni}, Y., {Di Matteo}, T., {Bird}, S., {et~al.} 2022, \mnras, 513, 670

\bibitem[{{Padoan} {et~al.}(2012){Padoan}, {Haugb{\o}lle}, \&
  {Nordlund}}]{padoan12}
{Padoan}, P., {Haugb{\o}lle}, T., \& {Nordlund}, {\r{A}}. 2012, \apj, 759, L27

\bibitem[{{Pezzulli} {et~al.}(2016){Pezzulli}, {Valiante}, \&
  {Schneider}}]{pezzulli16}
{Pezzulli}, E., {Valiante}, R., \& {Schneider}, R. 2016, \mnras, 458, 3047

\bibitem[{{Pfister} {et~al.}(2019){Pfister}, {Volonteri}, {Dubois}, {Dotti}, \&
  {Colpi}}]{pfister19}
{Pfister}, H., {Volonteri}, M., {Dubois}, Y., {Dotti}, M., \& {Colpi}, M. 2019,
  \mnras, 486, 101

\bibitem[{{Planck Collaboration} {et~al.}(2016){Planck Collaboration}, {Ade, P.
  A. R.}, {Aghanim, N.}, {Arnaud, M.}, {Ashdown, M.}, {Aumont, J.},
  {Baccigalupi, C.}, {Banday, A. J.}, {Barreiro, R. B.}, {Bartlett, J. G.},
  {Bartolo, N.}, {Battaner, E.}, {Battye, R.}, {Benabed, K.}, {Beno{\^\i}t,
  A.}, {Benoit-L{\'e}vy, A.}, {Bernard, J.-P.}, {Bersanelli, M.}, {Bielewicz,
  P.}, {Bock, J. J.}, {Bonaldi, A.}, {Bonavera, L.}, {Bond, J. R.}, {Borrill,
  J.}, {Bouchet, F. R.}, {Boulanger, F.}, {Bucher, M.}, {Burigana, C.},
  {Butler, R. C.}, {Calabrese, E.}, {Cardoso, J.-F.}, {Catalano, A.},
  {Challinor, A.}, {Chamballu, A.}, {Chary, R.-R.}, {Chiang, H. C.}, {Chluba,
  J.}, {Christensen, P. R.}, {Church, S.}, {Clements, D. L.}, {Colombi, S.},
  {Colombo, L. P. L.}, {Combet, C.}, {Coulais, A.}, {Crill, B. P.}, {Curto,
  A.}, {Cuttaia, F.}, {Danese, L.}, {Davies, R. D.}, {Davis, R. J.}, {de
  Bernardis, P.}, {de Rosa, A.}, {de Zotti, G.}, {Delabrouille, J.},
  {D{\'e}sert, F.-X.}, {Di Valentino, E.}, {Dickinson, C.}, {Diego, J. M.},
  {Dolag, K.}, {Dole, H.}, {Donzelli, S.}, {Dor{\'e}, O.}, {Douspis, M.},
  {Ducout, A.}, {Dunkley, J.}, {Dupac, X.}, {Efstathiou, G.}, {Elsner, F.},
  {En{\ss}lin, T. A.}, {Eriksen, H. K.}, {Farhang, M.}, {Fergusson, J.},
  {Finelli, F.}, {Forni, O.}, {Frailis, M.}, {Fraisse, A. A.}, {Franceschi,
  E.}, {Frejsel, A.}, {Galeotta, S.}, {Galli, S.}, {Ganga, K.}, {Gauthier, C.},
  {Gerbino, M.}, {Ghosh, T.}, {Giard, M.}, {Giraud-H{\'e}raud, Y.}, {Giusarma,
  E.}, {Gjerl{\o}w, E.}, {Gonz{\'a}lez-Nuevo, J.}, {G{\'o}rski, K. M.},
  {Gratton, S.}, {Gregorio, A.}, {Gruppuso, A.}, {Gudmundsson, J. E.}, {Hamann,
  J.}, {Hansen, F. K.}, {Hanson, D.}, {Harrison, D. L.}, {Helou, G.},
  {Henrot-Versill{\'e}, S.}, {Hern{\'a}ndez-Monteagudo, C.}, {Herranz, D.},
  {Hildebrandt, S. R.}, {Hivon, E.}, {Hobson, M.}, {Holmes, W. A.}, {Hornstrup,
  A.}, {Hovest, W.}, {Huang, Z.}, {Huffenberger, K. M.}, {Hurier, G.}, {Jaffe,
  A. H.}, {Jaffe, T. R.}, {Jones, W. C.}, {Juvela, M.}, {Keih{\"a}nen, E.},
  {Keskitalo, R.}, {Kisner, T. S.}, {Kneissl, R.}, {Knoche, J.}, {Knox, L.},
  {Kunz, M.}, {Kurki-Suonio, H.}, {Lagache, G.}, {L{\"a}hteenm{\"a}ki, A.},
  {Lamarre, J.-M.}, {Lasenby, A.}, {Lattanzi, M.}, {Lawrence, C. R.}, {Leahy,
  J. P.}, {Leonardi, R.}, {Lesgourgues, J.}, {Levrier, F.}, {Lewis, A.},
  {Liguori, M.}, {Lilje, P. B.}, {Linden-V{\o}rnle, M.}, {L{\'o}pez-Caniego,
  M.}, {Lubin, P. M.}, {Mac{\'\i}as-P{\'e}rez, J. F.}, {Maggio, G.}, {Maino,
  D.}, {Mandolesi, N.}, {Mangilli, A.}, {Marchini, A.}, {Maris, M.}, {Martin,
  P. G.}, {Martinelli, M.}, {Mart{\'\i}nez-Gonz{\'a}lez, E.}, {Masi, S.},
  {Matarrese, S.}, {McGehee, P.}, {Meinhold, P. R.}, {Melchiorri, A.}, {Melin,
  J.-B.}, {Mendes, L.}, {Mennella, A.}, {Migliaccio, M.}, {Millea, M.}, {Mitra,
  S.}, {Miville-Desch{\^e}nes, M.-A.}, {Moneti, A.}, {Montier, L.}, {Morgante,
  G.}, {Mortlock, D.}, {Moss, A.}, {Munshi, D.}, {Murphy, J. A.}, {Naselsky,
  P.}, {Nati, F.}, {Natoli, P.}, {Netterfield, C. B.}, {N{\o}rgaard-Nielsen, H.
  U.}, {Noviello, F.}, {Novikov, D.}, {Novikov, I.}, {Oxborrow, C. A.}, {Paci,
  F.}, {Pagano, L.}, {Pajot, F.}, {Paladini, R.}, {Paoletti, D.}, {Partridge,
  B.}, {Pasian, F.}, {Patanchon, G.}, {Pearson, T. J.}, {Perdereau, O.},
  {Perotto, L.}, {Perrotta, F.}, {Pettorino, V.}, {Piacentini, F.}, {Piat, M.},
  {Pierpaoli, E.}, {Pietrobon, D.}, {Plaszczynski, S.}, {Pointecouteau, E.},
  {Polenta, G.}, {Popa, L.}, {Pratt, G. W.}, {Pr{\'e}zeau, G.}, {Prunet, S.},
  {Puget, J.-L.}, {Rachen, J. P.}, {Reach, W. T.}, {Rebolo, R.}, {Reinecke,
  M.}, {Remazeilles, M.}, {Renault, C.}, {Renzi, A.}, {Ristorcelli, I.},
  {Rocha, G.}, {Rosset, C.}, {Rossetti, M.}, {Roudier, G.}, {Rouill{\'e}
  d'Orfeuil, B.}, {Rowan-Robinson, M.}, {Rubi{\~n}o-Mart{\'\i}n, J. A.},
  {Rusholme, B.}, {Said, N.}, {Salvatelli, V.}, {Salvati, L.}, {Sandri, M.},
  {Santos, D.}, {Savelainen, M.}, {Savini, G.}, {Scott, D.}, {Seiffert, M. D.},
  {Serra, P.}, {Shellard, E. P. S.}, {Spencer, L. D.}, {Spinelli, M.},
  {Stolyarov, V.}, {Stompor, R.}, {Sudiwala, R.}, {Sunyaev, R.}, {Sutton, D.},
  {Suur-Uski, A.-S.}, {Sygnet, J.-F.}, {Tauber, J. A.}, {Terenzi, L.},
  {Toffolatti, L.}, {Tomasi, M.}, {Tristram, M.}, {Trombetti, T.}, {Tucci, M.},
  {Tuovinen, J.}, {T{\"u}rler, M.}, {Umana, G.}, {Valenziano, L.}, {Valiviita,
  J.}, {Van Tent, F.}, {Vielva, P.}, {Villa, F.}, {Wade, L. A.}, {Wandelt, B.
  D.}, {Wehus, I. K.}, {White, M.}, {White, S. D. M.}, {Wilkinson, A.}, {Yvon,
  D.}, {Zacchei, A.}, \& {Zonca, A.}}]{planck16}
{Planck Collaboration}, {Ade, P. A. R.}, {Aghanim, N.}, {et~al.} 2016, A\&A,
  594, A13

\bibitem[{{Regan} {et~al.}(2019){Regan}, {Downes}, {Volonteri}, {Beckmann},
  {Lupi}, {Trebitsch}, \& {Dubois}}]{regan19}
{Regan}, J.~A., {Downes}, T.~P., {Volonteri}, M., {et~al.} 2019, \mnras, 486,
  3892

\bibitem[{{Reines} \& {Volonteri}(2015)}]{reines15}
{Reines}, A.~E. \& {Volonteri}, M. 2015, \apj, 813, 82

\bibitem[{{Rennehan} {et~al.}(2023){Rennehan}, {Babul}, {Moa}, \&
  {Dav{\'e}}}]{rennehan23}
{Rennehan}, D., {Babul}, A., {Moa}, B., \& {Dav{\'e}}, R. 2023, arXiv e-prints,
  arXiv:2309.15898

\bibitem[{{Ricarte} {et~al.}(2023){Ricarte}, {Narayan}, \& {Curd}}]{ricarte23}
{Ricarte}, A., {Narayan}, R., \& {Curd}, B. 2023, \apjl, 954, L22

\bibitem[{{Sala} {et~al.}(2021){Sala}, {Cenci}, {Capelo}, {Lupi}, \&
  {Dotti}}]{sala21}
{Sala}, L., {Cenci}, E., {Capelo}, P.~R., {Lupi}, A., \& {Dotti}, M. 2021,
  \mnras, 500, 4788

\bibitem[{{Sassano} {et~al.}(2023){Sassano}, {Capelo}, {Mayer}, {Schneider}, \&
  {Valiante}}]{sassano23}
{Sassano}, F., {Capelo}, P.~R., {Mayer}, L., {Schneider}, R., \& {Valiante}, R.
  2023, \mnras, 519, 1837

\bibitem[{{S{\c a}dowski} {et~al.}(2016){S{\c a}dowski}, {Lasota},
  {Abramowicz}, \& {Narayan}}]{sadowski16}
{S{\c a}dowski}, A., {Lasota}, J.-P., {Abramowicz}, M.~A., \& {Narayan}, R.
  2016, \mnras, 456, 3915

\bibitem[{{Schauer} {et~al.}(2017){Schauer}, {Regan}, {Glover}, \&
  {Klessen}}]{schauer17}
{Schauer}, A. T.~P., {Regan}, J., {Glover}, S. C.~O., \& {Klessen}, R.~S. 2017,
  \mnras, 471, 4878

\bibitem[{{Schaye} {et~al.}(2015){Schaye}, {Crain}, {Bower}, {Furlong},
  {Schaller}, {Theuns}, {Dalla Vecchia}, {Frenk}, {McCarthy}, {Helly},
  {Jenkins}, {Rosas-Guevara}, {White}, {Baes}, {Booth}, {Camps}, {Navarro},
  {Qu}, {Rahmati}, {Sawala}, {Thomas}, \& {Trayford}}]{schaye15}
{Schaye}, J., {Crain}, R.~A., {Bower}, R.~G., {et~al.} 2015, \mnras, 446, 521

\bibitem[{{Shakura} \& {Sunyaev}(1973)}]{shakura73}
{Shakura}, N.~I. \& {Sunyaev}, R.~A. 1973, \aap, 24, 337

\bibitem[{{S{\k{a}}dowski} {et~al.}(2016){S{\k{a}}dowski}, {Lasota},
  {Abramowicz}, \& {Narayan}}]{sadowski16b}
{S{\k{a}}dowski}, A., {Lasota}, J.-P., {Abramowicz}, M.~A., \& {Narayan}, R.
  2016, \mnras, 456, 3915

\bibitem[{{S{\k{a}}dowski} \& {Narayan}(2016)}]{sadowski16a}
{S{\k{a}}dowski}, A. \& {Narayan}, R. 2016, \mnras, 456, 3929

\bibitem[{{Soltan}(1982)}]{soltan82}
{Soltan}, A. 1982, \mnras, 200, 115

\bibitem[{{Springel}(2005)}]{springel05}
{Springel}, V. 2005, \mnras, 364, 1105

\bibitem[{{Springel} {et~al.}(2005){Springel}, {Di Matteo}, \&
  {Hernquist}}]{springel05a}
{Springel}, V., {Di Matteo}, T., \& {Hernquist}, L. 2005, \mnras, 361, 776

\bibitem[{{Springel} {et~al.}(2008){Springel}, {Wang}, {Vogelsberger},
  {Ludlow}, {Jenkins}, {Helmi}, {Navarro}, {Frenk}, \& {White}}]{springel08}
{Springel}, V., {Wang}, J., {Vogelsberger}, M., {et~al.} 2008, \mnras, 391,
  1685

\bibitem[{{Stone} {et~al.}(2023){Stone}, {Lyu}, {Rieke}, \&
  {Alberts}}]{stone23a}
{Stone}, M.~A., {Lyu}, J., {Rieke}, G.~H., \& {Alberts}, S. 2023, \apj, 953,
  180

\bibitem[{{Stone} {et~al.}(2024){Stone}, {Lyu}, {Rieke}, {Alberts}, \&
  {Hainline}}]{stone24}
{Stone}, M.~A., {Lyu}, J., {Rieke}, G.~H., {Alberts}, S., \& {Hainline}, K.~N.
  2024, \apj, 964, 90

\bibitem[{{Tanaka} \& {Haiman}(2009)}]{tanaka09}
{Tanaka}, T. \& {Haiman}, Z. 2009, \apj, 696, 1798

\bibitem[{{Tchekhovskoy} \& {Giannios}(2015)}]{tchekhovskoy15}
{Tchekhovskoy}, A. \& {Giannios}, D. 2015, \mnras, 447, 327

\bibitem[{{Tenneti} {et~al.}(2018){Tenneti}, {Di Matteo}, {Croft}, {Garcia}, \&
  {Feng}}]{tenneti18}
{Tenneti}, A., {Di Matteo}, T., {Croft}, R., {Garcia}, T., \& {Feng}, Y. 2018,
  \mnras, 474, 597

\bibitem[{{Tremmel} {et~al.}(2015){Tremmel}, {Governato}, {Volonteri}, \&
  {Quinn}}]{tremmel15}
{Tremmel}, M., {Governato}, F., {Volonteri}, M., \& {Quinn}, T.~R. 2015,
  \mnras, 451, 1868

\bibitem[{{Tremmel} {et~al.}(2017){Tremmel}, {Karcher}, {Governato},
  {Volonteri}, {Quinn}, {Pontzen}, {Anderson}, \& {Bellovary}}]{tremmel17}
{Tremmel}, M., {Karcher}, M., {Governato}, F., {et~al.} 2017, \mnras, 470, 1121

\bibitem[{{Uchiyama} {et~al.}(2018){Uchiyama}, {Toshikawa}, {Kashikawa},
  {Overzier}, {Chiang}, {Marinello}, {Tanaka}, {Niino}, {Ishikawa}, {Onoue},
  {Ichikawa}, {Akiyama}, {Coupon}, {Harikane}, {Imanishi}, {Kodama},
  {Komiyama}, {Lee}, {Lin}, {Miyazaki}, {Nagao}, {Nishizawa}, {Ono}, {Ouchi},
  \& {Wang}}]{uchiyama18}
{Uchiyama}, H., {Toshikawa}, J., {Kashikawa}, N., {et~al.} 2018, \pasj, 70, S32

\bibitem[{{Volonteri} {et~al.}(2021){Volonteri}, {Habouzit}, \&
  {Colpi}}]{Volonteri21}
{Volonteri}, M., {Habouzit}, M., \& {Colpi}, M. 2021, Nature Reviews Physics,
  3, 732

\bibitem[{{Volonteri} {et~al.}(2015){Volonteri}, {Silk}, \&
  {Dubus}}]{volonteri15}
{Volonteri}, M., {Silk}, J., \& {Dubus}, G. 2015, \apj, 804, 148

\bibitem[{{Xie} \& {Yuan}(2012)}]{xie12}
{Xie}, F.-G. \& {Yuan}, F. 2012, \mnras, 427, 1580

\bibitem[{{Yang} {et~al.}(2023){Yang}, {Wang}, {Fan}, {Hennawi}, {Barth},
  {Ba{\~n}ados}, {Sun}, {Liu}, {Cai}, {Jiang}, {Li}, {Onoue}, {Schindler},
  {Shen}, {Wu}, {Bhowmick}, {Bieri}, {Blecha}, {Bosman}, {Champagne}, {Colina},
  {Connor}, {Costa}, {Davies}, {Decarli}, {De Rosa}, {Drake}, {Egami},
  {Eilers}, {Evans}, {Farina}, {Habouzit}, {Haiman}, {Jin}, {Jun}, {Kakiichi},
  {Khusanova}, {Kulkarni}, {Loiacono}, {Lupi}, {Mazzucchelli}, {Pan},
  {Rojas-Ruiz}, {Strauss}, {Tee}, {Trakhtenbrot}, {Trebitsch}, {Venemans},
  {Vestergaard}, {Volonteri}, {Walter}, {Xie}, {Yue}, {Zhang}, {Zhang}, \&
  {Zou}}]{yang23}
{Yang}, J., {Wang}, F., {Fan}, X., {et~al.} 2023, \apjl, 951, L5

\bibitem[{{Yuan} \& {Narayan}(2014)}]{yuan14}
{Yuan}, F. \& {Narayan}, R. 2014, \araa, 52, 529

\bibitem[{{Yue} {et~al.}(2023){Yue}, {Eilers}, {Simcoe}, {Mackenzie},
  {Matthee}, {Kashino}, {Bordoloi}, {Lilly}, \& {Naidu}}]{yue23}
{Yue}, M., {Eilers}, A.-C., {Simcoe}, R.~A., {et~al.} 2023, arXiv e-prints,
  arXiv:2309.04614

\bibitem[{{Zhu} {et~al.}(2022){Zhu}, {Li}, {Li}, {Maji}, {Yajima}, {Schneider},
  \& {Hernquist}}]{zhu22}
{Zhu}, Q., {Li}, Y., {Li}, Y., {et~al.} 2022, \mnras, 514, 5583

\end{thebibliography}

\end{document}